\documentclass[a4paper,10pt,twocolumn]{article}
\usepackage[utf8]{inputenc}
\usepackage[affil-it]{authblk}
\usepackage{amsmath}
\usepackage{amssymb}
\usepackage[english]{babel}
\usepackage{geometry}
\usepackage{hyperref}
\usepackage{color}
\usepackage{graphicx}
\usepackage{subfig}
\usepackage{mwe}
\usepackage[font=small,labelfont=bf]{caption}

 \global\long\def\halfsize{0.9\columnwidth}

\begin{document}
\title{On the order reduction}
\author{Waleska P. F. de Medeiros \footnote{e-mail: waleska@fis.unb.br} and Daniel Müller \footnote{e-mail: muller@fis.unb.br}\\
Instituto de Física, Universidade de Brasília,  Brasília, Brazil}

\maketitle

{\small In this work we present an extension of the technique of the order reduction to higher perturbative approximations in an iterative fashion. The intention is also to analyze more carefully the conditions for the validity of the order reduction technique. With this in mind, a few simple situations in which the iterative order reduction converges analytically to the exact solutions are presented as examples. It is discovered that the order reduction as a perturbative iterative technique does not converge in the weak coupling limit as most of the known perturbative schemes, at least when applied to these examples. Also, considering these specific examples, the convergence of the order reduction occurs in strong coupling regimes. As a more realistic case, the order reduction is applied to Starobinsky's inflationary model is presented. It is verified that the method converges to the inflationary solution in the slow-roll regime.}
\section{Introduction} 

Higher derivatives (higher than second-order) are usually due to radiative corrected effective theories. They were investigated for the first time in the context of modifications in the equation of motion of a charged particle by Lorentz and Abraham \cite{Abraham:1905}. After that, this problem was studied again by Dirac \cite{dirac1938classical}, who found nonphysical solutions; runaway and pre-acceleration connected to the higher derivatives. Today, both the runaway and the pre-acceleration solutions can be resolved by considering a non-pointwise particle or using the first quantization \cite{Moniz:1976kr}. In  \cite{Faci2016} the runaway solution is addressed. Self-force is a subject of their own studied by many others, see, for instance, \cite{poisson2011motion}. 

In the context of the self-force, the order reduction was initially proposed by Landau, Lifshitz \cite{landau1951classical} which is defined only when there is an external source. It is expected that, in the presence of sources, it is possible to control the external frequencies such that they are always much smaller than the natural frequencies of the system with a good convergence of the order reduction. This scenario does not address the issue of transients. On the other hand, concerning effective theories especially gravitational, the vacuum case is more interesting as there is a common belief that, near the singularity, all other fields should become irrelevant. 

A slightly different approach is developed by Simon \cite{PhysRevD.41.3720}, \cite{PhysRevD.43.3308}, \cite{PhysRevD.45.1953}.
In Simon's construction, there are no external sources and yet it is still possible to control a time scale to be much smaller than the natural frequency of the system, which guarantees the convergence of order reduction. 

The present work intends to verify more carefully the conditions for the validity of the perturbative technique of order reduction, and also to apply it to the gravitational situation which is Starobinsky inflation \cite{Starobinsky:1980te}. According to the latest CMBR observations  \cite{refId0}, \cite{2019BAAS...51g.194H}, \cite{Collaboration2011} Starobinsky's model is the one that best fits the scalar-tensor ratio amplitude. Besides that, the theory has some transition from the ultraviolet to the infrared sector of gravity, which, avoiding the tachyon, presents the graceful exit from inflation. We must note that Starobinsky's inflationary solution was obtained earlier in Jordan frame by Ruzmaikina and Ruzmaikin \cite{1969JETP...30..372R}. Inflation models with plateau type potential reproduce Starobinsky's model, for further reading see, for instance, \cite{PhysRevD.98.083538}, \cite{Tenkanen:2020cvw} and \cite{Rubio_2019} about Higgs inflation, an inflationary model that a scalar field is coupled non-minimally to gravity and in \cite{Tenkanen:2020cvw} it is allowed to contain terms of $R^2$ in the action.

The model of effective gravity addressed in this context occurs in a scenario in which quantized fields are considered in a classical gravitational background, see, for instance, \cite{dewitt1965dynamical}. The counterterms necessary for a consistent theory 
\begin{equation}
    {
    S=\int \mathrm{d}^4x\sqrt{-g}\frac{m^2_p}{2}\left \{ R-\beta R^2 +\alpha\left[R_{\mu\nu}R^{\mu\nu}-\frac{1}{3}R^2 \right]\right \}}\label{ao}
\end{equation}
include Starobinsky's model.
It is well known that the value of $\beta\approx -1.305\times10^{9}$ $m_{p}^{-2}$ is set by CMBR observations \cite{refId0}, see also \cite{PhysRevD.98.083538}, \cite{Gorbunov_2015}.

To our knowledge, the first to apply the order reduction to effective gravity were Bel and Zia \cite{Bel:1985zz}, after that, of course, also by Simon \cite{PhysRevD.43.3308}, \cite{PhysRevD.45.1953} and Parker \cite{Parker:1993dk}; the method is very well accepted academically \cite{poisson2011motion}, \cite{gralla2009rigorous}.

The order reduction in the original setting has some deep differences between what is suggested in this present work, which we point out in the following. First, in the original setting, there is no attempt to obtain an iterative approach since the order reduction was meant to be used together with loop expansion of the effective action. In this setting, $2nd$ perturbative approximation would also require $2nd$ loop corrections resulting in a different gravitational theory. Following this reasoning, originally the $R^2$ term which is a mandatory term for $1st$ loop is treated perturbatively as compared to the Einstein term $R$. Also, originally the field equations where written for the metric scale factor. While the perturbative technique proposed hitherto is written for the gravitational theory \eqref{ao} with no intention to follow a loop expansion approximation. The terms $R^2$ and $R$ are thought of at the same footing and the Hubble parameter is chosen as variable instead of the scale factor.

Concerning both point particle self-force and effective gravity, there is a drastic difference among them which we must emphasize here. In the case of effective gravitational theories, it is not possible to know beforehand which are physically acceptable solutions. For example, solutions were found with no initial singularities \cite{Starobinsky:1980te} and also both with no initial singularity and with no particle horizons \cite{Anderson:1983nq}, \cite{Anderson:1984jf}, \cite{Anderson:1985ds}. Also, instabilities as tachyon were pointed out by \cite{PhysRevD.17.414} and \cite{MULLER:2014jaa} for the sign of the regularization parameters $\alpha < 0$ and $\beta>0$ in eq. (\ref{ao}). Usually \cite{PhysRevD.45.1953}, \cite{Starobinsky:1980te},  \cite{Parker:1993dk}, \cite{Anderson:1983nq}, \cite{Anderson:1984jf}, \cite{Anderson:1985ds}
 also consider, at the level of the equation of motion, an additional term with zero covariant divergence in all conformally flat models, which is due to Ginzburg \cite{Ginzburg1971}, \cite{Davies:1977ze}. In this present work, Ginzburg's term is not included. 

Instabilities in higher derivatives theories are known since Ostrogradsky's time \cite{Ostrogradsky}. As it can be seen, by following \cite{PhysRev.79.145}, this kind of instability is known today as ghosts \cite{Stelle:1977ry} characterized by kinetic terms with opposite signs. It must be stressed that it is not possible to eliminate the ghost by appropriate choices of the parameters of the theory, as is the case for the above-mentioned tachyon. Strictly classically speaking, as long as the perturbations remain sufficiently small, the degrees of freedom with different signs remain all free and this kind of instability does not show. Of course, this situation changes drastically in a quantized theory, where it is mandatory a stable vacuum. 

In this work, we learned that at least, when applied to the examples here, analyzed the order reduction as a perturbative iterative technique does not converge in the weak coupling limit, as most of the known perturbative schemes. Both cases with and without sources are investigated. Instead, the convergence of the order reduction occurs in strong coupling regimes. For a very interesting case of perturbative convergence in the strong coupling, see, for instance, the article of Bender and Wu \cite{1969PhRv..184.1231B}. 

The paper is organized as follows: in section \ref{sec1}, we present some particular situations in which the technique of the order reduction converges to the analytic solution. In this context, the technique is applied to the harmonic oscillator with and without source and to the ALD equation in the absence of gravitational fields with a constant electric field $E$ in the $x$ direction as a particular source. In the Appendix, it is shown the convergence of the recurrence relation uniquely to the overdamped solution of the harmonic oscillator in the absence of source. In section \ref{sec2}, the order reduction is applied to Starobinsky's inflationary model and compared to the exact numerical solution of the field equations in the slow-roll regime. Finally, section \ref{summary} contains a summary of the results and conclusions.

For numerical codes, we used gnu/gsl ode package, explicit embedded
Runge-Kutta Prince-Dormand (8, 9) method on Linux. The codes
were obtained using the algebraic manipulator Maple 16. The following conventions and units are taken $R^{\mu}_{\nu\sigma,\delta }=\Gamma ^{\mu}_{\nu\delta,\sigma}-...$, $R^{\mu}_{\nu}=R^{\sigma}_{\mu\sigma\nu}$, $R=R^{\mu}_{\mu}$, metric signature $-+++$, the greek indices $\mu$, $\nu$,.. run from $0-3$ and $G=\hbar=c=1$.

\section{Particular Situations}\label{sec1}
We will begin our discussion with the harmonic oscillator already in the stationary regime to emphasize the iterative technique of the order reduction 

\begin{equation}
\epsilon x''+x'+\frac{\omega ^{2}}{\gamma^2}x=\frac{f_{0}}{\gamma^2} e^{i({\Omega}/{\gamma}) \tau}. \label{dfho}
\end{equation}
Here $\omega$ is the natural frequency of the free system, $\gamma$ is the damping coefficient, $\Omega$ is the external frequency and where, as usual, $\epsilon$ is a dimensionless perturbative parameter which is set to unity in the end. The derivatives are with respect to the dimensionless time $\tau\equiv \gamma t$. The application of the order reduction consists in neglecting the higher derivatives terms, where in this case considering $|x^{\prime \prime}| < |x^\prime|<|x|,$ gives to lowest order
\begin{equation}
x'_{1}+\frac{\omega ^{2}}{\gamma^2}x_{1}=\frac{f_{0}}{\gamma^2}e^{i({\Omega}/{\gamma}) \tau},
\end{equation}
which can be easily solved assuming $x_1=c_{1}e^{i({\Omega}/{\gamma}) \tau}$, when $c_1$ is
\begin{equation}
{{c_{1}}=\frac{f_{0}}{\omega ^{2}+i\gamma \Omega }.} \label{c1}
\end{equation}
To second order, 
\begin{equation}
\epsilon x''_1+x'_{2}+\frac{\omega ^{2}}{\gamma^2}x_{2}=\frac{f_{0}}{\gamma^2}e^{i (\Omega/\gamma) \tau},
\end{equation}
assuming $x_{2}=c_{2}e^{i({\Omega}/{\gamma}) \tau}$ and using (\ref{c1}) results in
\begin{equation}
{{c_{2}}=\frac{f_{0}+\Omega ^{2}\epsilon c_1}{\omega ^{2}+i\gamma \Omega }.}
\end{equation}

Successively, 
\begin{equation}
{{c}_{n+1}=\frac{f_0+\Omega ^{2}\epsilon {c_{n}}}{\omega ^{2}+i\gamma \Omega },}
\end{equation}
and therefore, as long as $\Omega/\omega<1$, the order reduction, in this case, converges to the exact particular solution 
\begin{align}
 x_{n\rightarrow \infty}& =\frac{f_{0}e^{i({\Omega}/{\gamma}) \tau}}{\omega ^{2}+i\gamma \Omega -\Omega ^{2}\epsilon}
\end{align}
for the non homogeneous equation (\ref{dfho}). This is the case when there are $2$ frequencies, an external one $\Omega$ and the natural frequency of the free system $\omega$ with $\Omega/\omega<1$ and this is the situation in which it is very well known the convergence of the order reduction first written in Landau-Lifshitz book \cite{landau1951classical}.

Now we turn to the issue of the transients. As it is well known, this system has $3$ regimes. The underdamped, overdamped, and critically damped. The underdamped is the one that, in some sense, reproduces the perturbative techniques in most textbooks, for example, in quantum field theory, where the damping is due to the weak coupling between the other fields and the free system. When the system is underdamped $|x^{\prime \prime}|$ is of the same order of $|x|(\omega/\gamma)^2$. Thus, the order reduction technique does not apply to the underdamped regime. This is a very deep difference between the order reduction and ordinary perturbative schemes which has not been stressed before.

To apply the order reduction in the homogeneous equation version of (\ref{dfho}),
\begin{align}
x^\prime_{n+1}+\frac{\omega^2}{\gamma^2}x_{n+1}=-\epsilon x_n^{\prime\prime},
\label{ordem_n}
\end{align}
it is also necessary that $|x^{\prime \prime}| < |x^\prime|<|x|$. To lowest order
\begin{equation}
{x'_1+\frac{\omega ^{2}}{\gamma^2}x_{1}=0.} \label{firstorder}
\end{equation}
Substituting $x_{1}=c_1e^{\lambda_{1}\tau}$ into (\ref{firstorder}) gives $\lambda _{1}=-{\omega ^{2}}/{\gamma^2 }$ with 
\begin{align}
    x_1=c_1e^{-\omega^2\tau/\gamma^2}
    \label{x_1}.
\end{align}
To obtain the next order $x_1$ from \eqref{x_1} is replaced into 
\begin{equation}
{\epsilon x''_1+x'_{2}+\frac{\omega ^{2}}{\gamma^2}x_2=0},
\label{secondorder}
\end{equation}
with solution
\begin{equation}
x_{2}=\left(c_2-\epsilon\frac{\omega^4}{\gamma^4 }c_1\tau\right)\exp\left({-\frac{\omega ^{2}}{\gamma^2 }\tau}\right),
\end{equation}
shows an additional constant $c_2$. The appearance of additional constants is a direct consequence of this method since for each perturbative order a differential equation must be solved. That is the reason why we choose to emphasize this point. 

In this particular case, \eqref{secondorder} is a first derivative ODE, and for each higher order perturbative approximation, there must be one additional constant. Since higher perturbative orders of whichever perturbative technique must contain the lower order approximations, these additional constants are uniquely determined and made equal to $c$. 

Then, written up to order order $5$ in $\epsilon$ the technique results in
\begin{align}
x= 
c\exp \left[-\left({\frac{\omega ^{2}}{\gamma^2 }}+\epsilon\frac{\omega^4}{\gamma^4 }+2\epsilon^2\frac{\omega^6}{\gamma^6}+5\epsilon^3\frac{\omega^{8}}{\gamma^8}+\right .\right.\nonumber\\\left.\left.+14\epsilon^4\frac{\omega^{10}}{\gamma ^{10}}+42\epsilon^5\frac{\omega^{12}}{\gamma ^{12}} \right )\tau\right].
\end{align} 
 
Facing \eqref{ordem_n} as a map, in the Appendix, it is discussed that this map is a contraction, and that the method converges to the exact solution 
\begin{align}
x=c\exp\left[\left(-1+\sqrt{1-4\epsilon\omega^2/\gamma^2} \right)\frac{\tau}{2\epsilon}\right],
\label{sol_exata}
\end{align}
which is a fixed point for this map. 

In this same Appendix, it is also discussed that the other fixed point, namely 
\begin{align}
x=c\exp\left[\left(-1-\sqrt{1-4\epsilon\omega^2/\gamma^2} \right)\frac{\tau}{2\epsilon}\right],
\label{2ponto_fixo}
\end{align}
it's not defined when $\epsilon\rightarrow 0$. While when $\epsilon\rightarrow 0$; both \eqref{sol_exata} has a well defined limit $x=ce^{-\omega^2\tau/\gamma^2}$, which coincides with the exact solution $x=ce^{-\omega^2\tau/\gamma^2}$ of \eqref{ordem_n}. This second fixed point, \eqref{2ponto_fixo}, then must be excluded and when $\epsilon\rightarrow 1$ we are left with the unique solution of \eqref{ordem_n} 
\begin{align*}
    x=c\exp\left[\left(-1+\sqrt{1-4\omega^2/\gamma^2} \right)\frac{\tau}{2}\right],
\end{align*}
led to conclude that this iterative procedure converges to this solution.

In the context of the charged particle, the relativistic Abraham-Lorentz-Dirac (ALD) equation without the presence of gravitational fields becomes  \cite{rohrlich2007classical}
\begin{equation}
{a^{\mu}=\frac{q}{m}{F^{\mu}}_{\nu}u^{\nu}+\frac{2}{3}\frac{q^{2}}{mc^{3}}\left ( {g^{\mu}}_{\nu}+u^{\mu}u_{\nu} \right )\dot{a}^{\nu}}, \label{ALD}
\end{equation}
where $q$ and $m$ are the charge and the mass of the particle, respectively, and $c$ is the the speed of light. The second term on the right hand side is the self-force on the charge resulting from its own electromagnetic field. 
 Let's proceed with the order reduction applied to (\ref{ALD}) for a constant electric field $E$ in the $x$ direction. Of course, this fulfils its convergence requirements, since the external source has zero frequency, which is always less than the natural frequencies of the system. To a first approximation, neglecting the highest order term, we have 
\begin{equation}
{a^{\mu}_{1}=\frac{q}{m}{F^{\mu}}_{\nu}u_1^{\nu}}.
\label{ordem1}
\end{equation}
Consider a time-like $u^\mu=(u^0,u^1, 0, 0)$ with $u_\mu u^\mu=-1$. Then, proceeding as described above for the oscillator, it is not difficult to find that the first-order solution (\ref{ordem1}) is $u^\mu_1=(\cosh(qEt/m),\sinh(qEt/m),0,0)$. The derivative of (\ref{ordem1}) substituted into the second term in the right-hand side of (\ref{ALD}) vanishes, showing that perturbatively the exact result is consistently obtained with the order reduction. This is not anything new, since this exact solution was found by Dirac himself \cite{dirac1938classical}, \cite{rohrlich2007classical}.\footnote{ It is very well known that $x^{\mu}=\left ( \xi \sinh\tau ,\xi \cosh\tau ,y,z \right )$ are Rindler coordinates \cite{rindler:1966}, and constant $\xi$ describe families of constant proper acceleration. } 

\section{Starobinsky Inflation}\label{sec2}
We will now apply the order reduction to Starobinsky's inflationary model \cite{Starobinsky:1980te}. As already mentioned in the introduction, this inflationary model is the one that best fits the scalar-tensor ratio amplitude, according to the latest CMBR observations  \cite{refId0}, \cite{2019BAAS...51g.194H}, \cite{Collaboration2011}. In this work, only the Jordan frame is chosen, and the model can be thought of as an effective action truncated at second order in field products (\ref{ao}).
Metric variations in (\ref{ao}) result in field equations of order-4. For the homogeneous isotropic line element $g_{ab}=\text{diag}[-1,e^{2a},e^{2a},e^{2a}]$ with zero spatial curvature there is the $00$ 
\begin{equation}
{\frac{1}{6}H^2+\beta \left [ -2\ddot{H}H-6\dot{H}H^2+\dot{H}^2 \right ]=0}\label{eq10},
\end{equation}
and the $11$
\begin{equation}
-\frac{1}{2}H^2-\frac{1}{3}\dot{H}+\beta \left [ 2\dddot{H}+12\ddot{H}H +9\dot{H}^2+18\dot{H}H^2\right ]=0\label{11},
\end{equation}
 equations of motion, where $H=\dot{a}$ is the Hubble parameter. Since the metric is isotropic, the terms that multiply $\alpha$ are canceled.
 The $00$ equation of motion is a constraint which is dynamically preserved, see for instance \cite{alma9934409023505154}, and it is used as a numerical check.

Again, the order reduction is applied. The $00$ equation of motion, results in the recurrence relation
\begin{equation}
{-2 \epsilon\beta\frac{\ddot{H}_n}{H_n}+\epsilon\beta\frac{\dot{H}^{2}_n}{H^2_n}+\frac{1}{6}\left [ 1-36\beta \dot{H} _{n+1}\right ]=0}, \label{00}
\end{equation}
where the parameter $\epsilon$ as before is dimensionless and set to unity in the end with $H_n\neq 0$ and the equation is dimensionless in the proper time t. The conditions used are $|\beta\ddot{H}|\ll  |H|$ and $|\beta\dot{H}^2|\ll  |H^2|$. The first and second slow-roll conditions for inflation are given by $|\ddot{H}|\ll  |\dot{H}H|$ and $|\dot{H}|\ll  |H^2|$, and there is some overlap in convergence region of the order reduction and slow-roll conditions. We mention that it is possible to rewrite the order reduction technique, \eqref{00}, in such a manner that its convergence conditions are identical with the slow-roll conditions.  

To first order, we have the solution of Ruzmaikina and Ruzmaikin \cite{1969JETP...30..372R}, which describes the slow-roll regime
\begin{equation}
{1-36 \beta\dot{H}_{1}=0\Rightarrow H_{1}=\frac{1}{36\beta}(t-t_0)},\label{13}
\end{equation}

To second order,
\begin{equation}
{H_{2}=\frac{1}{36\beta}(t-t_0)-\frac{\epsilon}{6(t-t_0)}, \label{14}
}
\end{equation}

To third order,
\begin{equation}
    H_3=\frac{1}{36\beta}(t-t_0)-\frac{\epsilon}{6(t-t_0)}-{\frac {8\beta\epsilon^2}{3(t-t_0)^{3}}},
\label{o3}
\end{equation}
 where in both cases $t_0$ is a constant of integration.

The method can be repeated for the next orders. 

For qualitative analysis, $\beta=-10$ is chosen in our numerical results. 

 \begin{figure}[h!]
        \centering
        \resizebox{\halfsize}{!}{\includegraphics{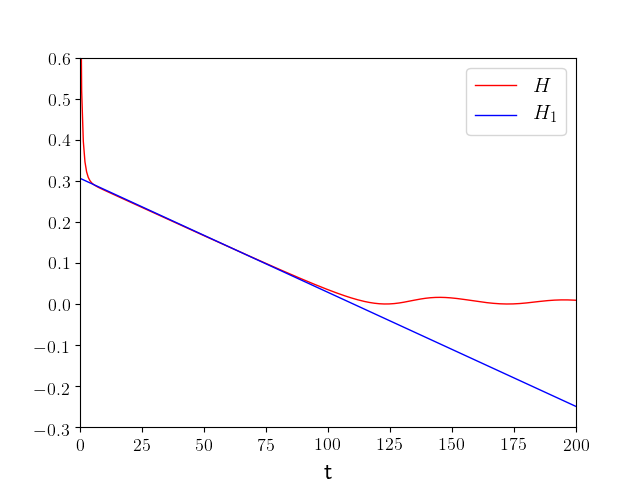}}
      \captionof{figure}{It is shown in red the exact numerical solution of eq. (\ref{eq10}) for $\beta=-10$, $H(t)$. Plotted in blue is the perturbative solution (\ref{13}), $H_1(t)$. The initial condition is chosen as $H(0)=1$, $\dot{H}(0)=-1.66666$, while the constant $t_0=110.2$ in (\ref{13}) is fixed by best fitting. It can be seen that the perturbative solution does not agree with the field equation towards the singularity for decreasing time. Both solutions also show disagreement in the linearized weak-field regime. On the other hand, both solutions show very good agreement in the slow-roll regime.
      }
      \label{fig1}
      \end{figure} 

        \begin{figure}[h!]
        \centering
        \begin{tabular}{c c}
    a) \resizebox{\halfsize}{!}{\includegraphics{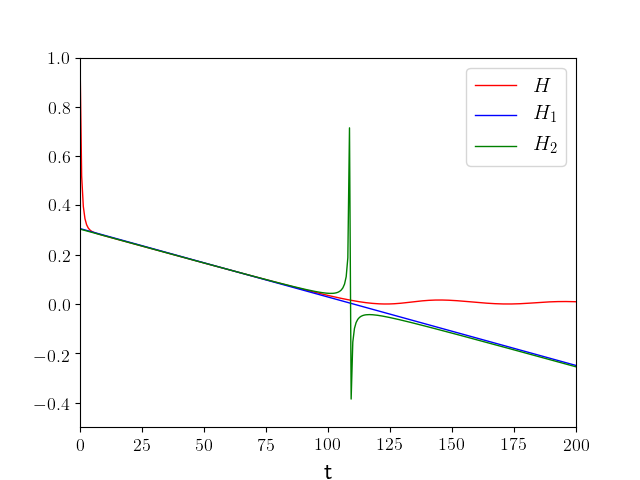}} & \\
    b) \resizebox{\halfsize}{!}{\includegraphics{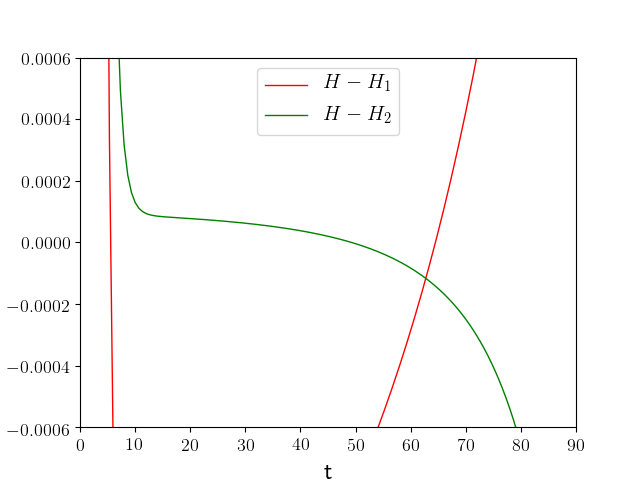}} \\ & 
        \end{tabular}
        \captionof{figure}{a) It is show in red the exact numeric solution of eq. (\ref{eq10}) for $\beta=-10$, $H(t)$. The perturbative approximations are plotted in blue for the eq. (\ref{13}), $H_1(t)$ and in green for the eq. (\ref{14}), $H_2(t)$. The initial condition is chosen as $H(0)=1$, $\dot{H}(0)=-1.66666$, while the constant $t_0=110.2$ in (\ref{13}) and $t_0=108.9$ in (\ref{14}) are fixed by best fitting.  b) The graph in red shows the difference between the exact numeric solution (\ref{eq10}), $H(t)$ and the analytical approximation (\ref{13}), $H_1(t)$. Plotted in green is the difference between the exact numeric solution (\ref{eq10}), $H(t)$ and the analytical approximation (\ref{14}), $H_2(t)$. It is possible to see a weak convergence for higher orders to the exact numerical solution in the regime of slow-roll.}\label{fig2} 
      \end{figure} 

 It is well-known that $H(t)$ decreases linearly (slow-roll or Ruzmaikina's regime) approaches zero and enters into the phase of the damped oscillations (reheating regime) \cite{Suen:1987gu}. See Figure \ref{fig1} and Figure \ref{fig2} panel a) in red the exact numeric solution of equation (\ref{eq10}) for $\beta=-10$, $H(t)$. It can be seen that the perturbative solutions do not agree with the field equation (\ref{eq10}) towards the singularity for decreasing time. Both solutions also show disagreement with the exact numeric solution in the linearized weak-field regime. This is expected, as both regions, named towards singularity and weak-field, do not fulfill the requirements for the order reduction. 
 
 On the other hand, both solutions show very good agreement in the slow-roll regime, as can be seen in Figure \ref{fig1} and Figure \ref{fig2} panel a). This is strongly connected to the choice of the constant $t=t_0$. For an inconvenient $t_0$, there will be no agreement whatsoever between the direct numeric solution and the order reduction method.
 
As shown in \eqref{o3}, in this case, the order reduction results in a Laurent series with non zero principal part with infinite terms. As it's well known this series will not converge in the limit $t\rightarrow t_0$  \cite{morse1953methods}, shown by the asymptotes in Figure \ref{fig1} and Figure \ref{fig2} panel a). The location of the asymptote in the weak-field limit of small oscillations is a consequence of the choice of the constant $t=t_0$ done exclusively to best fit the method in Ruzmaikina's regime. 
 For higher orders, the asymptotes appear alternated in pairs due to successive powers of $\beta$, which must be negative to avoid the tachyon, as mentioned in the introduction.
 
 Besides that, it is possible to see that higher orders of the order reduction method show some convergence to the exact numeric solution as shown in Figure \ref{fig2} panel b). It must also be mentioned that the convergence of the order reduction is slow. 

The situation changes in the presence of sources or spatial curvature since then the field equations will depend explicitly in the scale factor.\footnote{ This discussion is strictly for spatially homogeneous space times.} For instance, in the presence of perfect fluid source $p=w\rho$ with EOS parameter $w$ the covariant conservation of this source $\nabla^bT_{ab}=0$ implies a dependence on the scale factor, $e^a$, $\rho=\rho_0(e^{a_0}/e^a)^{3(w+1)}$. 
In this case, the method will necessarily present second time derivatives to lowest order instead of first time derivatives in $H$. If the source is also considered perturbatively, to lowest order the first derivative equation \eqref{13} is replaced by 
 \begin{align}
     1-36\beta\ddot{a}_1=0,
     \label{eq_a}
 \end{align}
remind that $H=\dot{a}$. The explicit dependence of the field equation on the scale factor, through the source $\rho=\rho_0(e^{a_0}/e^a)^{3(w+1)}$ will come in higher perturbative approximations, by assumption.
 
We end this section by briefly addressing the choice of variables as compared to Simon and Parker's work. The reason for this is that this order reduction which we are presenting here is very sensitive to the choice of the lowest perturbative approximation. 
As already mentioned above, the field equations written with respect to the scale factor will have an additional time derivative as compared to the same field equations written with respect to $H$ as it's done in this present work. The lowest order system in the order reduction must be chosen in accordance to which regime of the solution is going to be reproduced by the method. If higher than first time derivatives of the scale factor are neglected in the lowest perturbative approximation of the order reduction, the method should present good agreement with linear growth of scale factor, $e^a \propto t$. On the other hand, keeping just first time derivatives of the scale factor to lowest order, Ruzmaikina's regime $H=\dot{a}\propto -t$ is not reproduced. 

There's another deep difference between Simon and Parker's work and this present one. In their work, the field equations are written with respect to the scale factor, and only first time derivatives are taken into account. As mentioned in the introduction, in their method there's no attempt to higher perturbative approximations, as we do here.
 
For a spatially flat homogeneous and isotropic space-time and a conformally invariant free quantum field in the conformal vacuum state, the expectation value of the energy momentum tensor of the quantum field will depend only on H and it's derivatives \cite{birrell_davies_1982} so that the technique presented here can be applied.

\section{Conclusion}\label{summary}

In this work, it is presented a simple extension of the order reduction technique to higher perturbative orders as an iterative technique. The analytical approximations following this technique are also compared with direct numerical evaluation of the equation of motion. First, we remark some considerations on the order reduction as follows.

In section \ref{sec1} a few examples are shown for which the technique converges to the exact solution, also to gain intuition. Surprisingly, the order reduction presents a very good agreement in strong coupling regimes. While in the weak coupling, it is inapplicable. Both situations with or without a source are analyzed. And we discovered that, without an external source, the technique only applies and converges to the non-oscillating solution which slowly approaches equilibrium. Remind that the weak coupling regime is excluded by the order reduction. While the case with external source falls into the class of problems mentioned in the introduction. It is possible to control the external frequency or a time scale to be much smaller than the natural frequency of the system and order reduction converges to the expected solution. As an example of perturbations in the strong coupling, see, for instance, the very interesting article of Bender and Wu \cite{1969PhRv..184.1231B}.

Also in \ref{sec1}, the order reduction is applied to the relativistic self-force problem in the absence of gravitational fields. It is considered a constant electric field $E$ in the $x$ direction as a particular source and the method gives the well known Rindler motion for the point charge. It must be mentioned that in this situation there is strong coupling and also there is an external time scale that is always much larger (a constant electric field) than the natural internal time scales. 

Previous applications of the order reduction to effective gravity seemed to be done only with the presence of sources \cite{PhysRevD.41.3720}, \cite{PhysRevD.43.3308}, \cite{PhysRevD.45.1953}, \cite{Parker:1993dk}. 

In section \ref{sec2} the order reduction is applied to Starobinsky's inflationary model. This cosmological model follows from quadratic gravity with a homogeneous isotropic line element and zero spatial curvature in absence of classical sources, vacuum. The order reduction is applied to the equation of motion (\ref{eq10}) resulting in the recurrence relation (\ref{00}). The convergence region has some overlap with the first and second slow-roll conditions for inflation. 
This recurrence relation (\ref{00}) is used to obtain successive analytical approximations that are compared to the direct numerical solution of equation (\ref{11}). Equation (\ref{eq10}) is dynamically conserved and is used to numerically check the code. 

It can be seen in Figure \ref{fig1} and Figure \ref{fig2} panel a) that the perturbative solution does not agree with the field equation (\ref{eq10}) towards the singularity for decreasing time. Both solutions also show disagreement with the exact numeric solution in the linearized weak-field regime. 

The asymptote present in $H_2(t)$ in Figure \ref{fig2} panel a) is consequence of the choice of the constant $t=t_0$. This choice of $t_0$ is intentionally made to best fit the exact numeric solution with the perturbative approximation in the slow-roll regime. This asymptote occurs in the weak-field regime, where the technique of the order reduction does not work. For the following perturbative approximations, the asymptotes appear alternated due to successive powers of $\beta$, which must be negative to avoid the tachyon, as mentioned in the introduction.

Moreover, we verify the convergence of the technique of the order reduction, as shown in Figure \ref{fig2}) panel b). It is possible to see that successive approximations of the order reduction method show some convergence to the exact numeric solution. It must also be mentioned that this convergence is slow. 

It is well known that order reduced equations present fewer solutions \cite{bender1999advanced}. This was one of the intentions of the order reduction technique to select the ones that are physically relevant \cite{Bel:1985zz}, \cite{Parker:1993dk}. This present work is in agreement with this reasoning. For all solutions analyzed hitherto, the perturbative order reduction in its convergence region approaches the physical solutions. Anyway, we must emphasize that there could be physical solutions that will not be detected by order reduction. For example, the order reduction does not apply towards the singularity and also in the free field oscillations described in section \ref{sec2}. 

As mentioned in the introduction in this present work the $R^2$ term in the gravitational Lagrangian \eqref{ao} is not disregarded as compared to the $R$ term. As a result, Einstein field equation at lowest perturbative order $G_{ab}=\kappa T_{ab}$ is not reproduced. Instead, in section \ref{sec2}, we see that the order reduced solution in its lowest approximation already presents contributions from the radiative corrected gravity, $R^2$. 

Also, as discussed in section \ref{sec2} the field equations written with respect to the scale factor will have an additional time derivative as compared to the same field equations written with respect to $H$ as it's done here. If higher than first time derivatives of the scale factor are going to be disregarded in the lowest perturbative approximation of the order reduction, the method should present good agreement with linear growth of scale factor, $e^a \propto t$. On the other hand, keeping just first time derivatives of the scale factor to lowest order, Ruzmaikina's regime $H=\dot{a}\propto -t$ is not reproduced in lowest order.

In the presence of sources as mentioned at the end of section \ref{sec2}, the correct choice of variable should be logarithmic of scale factor instead of the Hubble parameter. The lowest order approximation should be given by \eqref{eq_a} and the source contribution should come in higher perturbative approximations, by assumption.

For a spatially flat homogeneous and isotropic space-time and a conformally invariant free quantum field in the conformal vacuum state, the expectation value of the energy momentum tensor of the quantum field will depend only on H and it's derivatives \cite{birrell_davies_1982} so that the technique presented here can be applied.

\section*{Appendix}
Consider the following first order differential equation for $x_{n+1}$ to be understood as an iteration map
\begin{align}
x^\prime_{n+1}+\frac{\omega^2}{\gamma^2}x_{n+1}=-\epsilon x^{\prime\prime}_n ,
\label{rel.rec}
\end{align}
where the parameter $0\leq \epsilon\leq 1$ and at the end is made $\epsilon=1$. We will check that the above iteration map is a contraction. Variables are changed to $x_n=e^{-\omega^2\tau/\gamma^2}y_n$ assuming that $y_n$ and its derivatives are limited functions in the time interval in question
\[
I=[\tau_0,\tau].
\]
Integrating by parts twice
\begin{align*}
&y^\prime_{n+1}=-\epsilon e^{\omega^2\tau/\gamma^2}x^{\prime\prime}_n,\\ &y_{n+1}-y_{n+1}^0=-\epsilon\int_{\tau_0}^\tau e^{\omega^2s/\gamma^2}x^{\prime\prime}_n(s)ds,\\
&y_{n+1}-y_{n+1}^0=-\epsilon \left[  e^{\omega^2s/\gamma^2}x^{\prime}_n(s) \right] _{\tau_0}^\tau+\epsilon\frac{\omega^2}{\gamma^2}\left[ e^{\omega^2s/\gamma^2}x_n(s)\right]_{\tau_0}^\tau\\
&-\epsilon\frac{\omega^4}{\gamma^4} \int_{\tau_0}^\tau e^{\omega^2s/\gamma^2}x_n(s)ds,\\
&y_{n+1}-y_{n+1}^0=-\epsilon \left[y^{\prime}_n(s) -2\frac{\omega^2}{\gamma^2}y_n(s)\right] _{\tau_0}^\tau-\epsilon\frac{\omega^4}{\gamma^4} \int_{\tau_0}^\tau y_n(s)ds,
\end{align*}
where $y_{n+1}^0$ is the initial condition. The metric is induced by the uniform norm 
\[
\|y(\tau)\| =\sup_{\tau\in I}|y(\tau)|.
\]
First, we show that for a given function $y$, its first iteration $y_1$ is within some upper limit, 
\[
\|y_1 -y_1^0\|<b, \;\;b>0.
\]
Begin with 
\begin{align}
&y_1-y_1^0=-\epsilon \left[y^{\prime}(s)-2\frac{\omega^2}{\gamma^2}y(s)\right] _{\tau_0}^\tau-\epsilon\frac{\omega^4}{\gamma^4} \int_{\tau_0}^\tau y(s)ds\label{demonstr.1}.
\end{align}
Then, considering that both  
\begin{align}
& |y(\tau)-y(\tau_0)|\leq \sup\left|\frac{dy}{d\tau}\right| \Delta \tau=
\|y^\prime \|\Delta \tau\nonumber\\
& |y^\prime(\tau)-y^\prime(\tau_0)|\leq \sup\left|\frac{d^2y}{d\tau^2}\right| \Delta \tau=\| y^{\prime \prime} \|\Delta \tau
\label{y-y0}
\end{align}
and that
\begin{align}
\left| \int_{\tau_0}^\tau y(s)ds\right|\leq \int_{\tau_0}^\tau|y(s)|ds\leq\| y\|\Delta\tau
\label{int_mod}
\end{align}
with $\Delta \tau=\tau-\tau_0 $ it is possible to rewrite $\|y_1-y^0 \|$ as 
\begin{align*}
    \|y_1-y_1^0 \|\leq \epsilon \left\{ \| y^{\prime\prime}\| +2\frac{\omega^2}{\gamma^2}\| y^{\prime}\|+\frac{\omega^4}{\gamma^4}\|y\| \right\}\Delta \tau.
\end{align*}
Since $y$ and its derivatives have definite norm, it is always possible to choose $\Delta \tau$ such that $\|y_1-y_1^0 \|<b$. 

Now, given two functions $y_a$ and $y_b$, we shall prove that 
 \begin{align*}
    \|y_a^1 -y_b^1\|\leq q\|y_a-y_b\|, 
\end{align*} 
for some  $0\leq q<1$. We will suppose that both $y^\prime$ and $y^{\prime\prime}$ have Lipschitz constants $L_1\geq 0$ and $L_2\geq 0$  in the time interval $I$ 
\begin{align*}
&\|y_a^\prime -y_b^\prime\|\leq L_1\|y_a-y_b\| & \|y_a^{\prime\prime} -y_b^{\prime\prime}\|\leq L_2\|y_a-y_b\|
\end{align*}
which is a rather strong condition, anyway reasonable, since by assumption, all these functions are limited in the considered time interval. Following \eqref{demonstr.1} for two distinct functions $y_a$ and $y_b$ with same initial condition $y_a^0=y_b^0$ and performing their difference results in
\begin{align*}
& y_a^1-y_b^1=-\epsilon \left[y^{\prime}_a(s)-y^{\prime}_b(s)-2\frac{\omega^2}{\gamma^2}\left( y_a(s)-y_b(s)\right)\right] _{\tau_0}^\tau\nonumber\\ &
-\epsilon\frac{\omega^4}{\gamma^4} \int_{\tau_0}^\tau(y_a(s)-y_b(s))ds.
\end{align*}
Keeping in mind \eqref{y-y0} and \eqref{int_mod} then 
\begin{align*}
&\| y_a^1-y_b^1\|\leq\epsilon\left\{ L_2+2\frac{\omega^2}{\gamma^2}L_1
+\frac{\omega^4}{\gamma^4} \right\}\|y_a -y_b\|\Delta\tau.
\end{align*}
It is always possible to choose a sufficiently small time interval $\Delta\tau$ such that the above relation is
\begin{align*}
&\| y_a^1-y_b^1\|\leq q\|y_a -y_b\|\
\end{align*}
with $0\leq q < 1$, which shows that the map is a contraction in the metric space of functions with uniform norm. Banach fixed point theorem states that, since it is a contraction map, it has a unique fixed point \cite{KreyszigErwin1989Ifaw}. 

It must be mentioned, since the analytic solution is known, that there are two fixed points $x_\infty^+$ and $x_\infty^-$ for \eqref{rel.rec} 
\begin{align*}
&x_\infty^\pm=c\exp\left[\left(-1\pm\sqrt{1-\frac{4\epsilon\omega^2}{\gamma^2}}\right) \frac{\tau}{2\epsilon}\right].
\end{align*}
On the other hand, it can be easily seen in \eqref{rel.rec} that when  $\epsilon\rightarrow 0$ the solution is  $x_n=e^{-\omega^2\tau/\gamma^2}$. Now, only one of the fixed points $x_\infty^\pm$ is consistent with this solution, namely
\begin{align*}
x_\infty^+=c\lim_{\epsilon=0}\left\{\exp\left[\left(-1+\sqrt{1-\frac{4\epsilon\omega^2}{\gamma^2}}\right) \frac{\tau}{2\epsilon}\right]\right\}=e^{-\omega^2\tau/\gamma^2}.
\end{align*}
 The other fixed point, $x_\infty^-$ does not have a well defined limit when $\epsilon\rightarrow 0$ and must be excluded.
 
 The iterative procedure \eqref{rel.rec}, when $\epsilon\rightarrow 1$, then converges to the unique solution 
 \[
 x=c\exp\left[\left(-1+\sqrt{1-\frac{4\omega^2}{\gamma^2}}\right) \frac{\tau}{2}\right].
 \]

 \section*{Acknowledgments}
We gratefully acknowledge Dra. R. F. P. Mendes, Dr. C. A. S. Maia, Dr. I. S. Ferreira, and Dr. A. Melikyan for corrections and improvements. We also thank the referee for the valuable suggestions and kind comments. W. P. F. de Medeiros wishes to thank the Brazilian agencies FAPDF and CAPES project number 88882.383677/2019-01. 

\bibliographystyle{utcaps}
\providecommand{\href}[2]{#2}\begingroup\raggedright\endgroup

\end{document}